\documentclass[final,2p,times,twocolumn]{elsarticle}

\usepackage{amssymb}
\usepackage{amsmath}

\usepackage{natbib}
\usepackage[colorlinks=false,linkcolor=black, citecolor=blue, urlcolor=blue]{hyperref}

\usepackage{url}
\usepackage{svg}
\usepackage{arydshln}
\usepackage{svg}
\usepackage{caption}
\usepackage{subcaption}
\usepackage{booktabs}
\usepackage{graphicx}

\journal{Information and Software Technology}

\begin{document}

\begin{frontmatter}

\title{Research Artifacts in Secondary Studies: A Systematic Mapping in Software Engineering}

\author[unihelsinki]{Aleksi Huotala}
\author[unidalhousie]{Miikka Kuutila}
\author[unihelsinki]{Mika Mäntylä}

\affiliation[unihelsinki]{organization={University of Helsinki},
             city={Helsinki},
             country={Finland}}
\affiliation[unidalhousie]{organization={Dalhousie University},
             city={Halifax},
             country={Canada}}


\begin{abstract}
\textbf{Context:} Systematic reviews (SRs) summarize state-of-the-art evidence in science, including software engineering (SE). 
\textbf{Objective:} Our objective is to evaluate how SRs report research artifacts and to provide a comprehensive list of these artifacts. 
\textbf{Method:} We examined 537 secondary studies published between 2013 and 2023 to analyze the availability and reporting of research artifacts.
\textbf{Results:} Our findings indicate that only 31.5\% of the reviewed studies include research artifacts. Encouragingly, the situation is gradually improving, as our regression analysis shows a significant increase in the availability of research artifacts over time. However, in 2023, just 62.0\% of secondary studies provide a research artifact while an even lower percentage, 30.4\% use a permanent repository with a digital object identifier (DOI) for storage.
\textbf{Conclusion:} To enhance transparency and reproducibility in SE research, we advocate for the mandatory publication of research artifacts in secondary studies.

\end{abstract}


\end{frontmatter}

\section{Introduction}

Systematic reviews (SRs) summarize state-of-the-art evidence in software engineering (SE). Artifact research in SE has previously focused on replicating experiments~\cite{cruz2019replication} and increasing reproducibility in technical areas of software engineering~\cite{heumuller2020publish}. SRs should include a research artifact with all necessary details for several reasons.

Reason 1: Replicability (Reproducibility or Repeatability). 
Replication is a hallmark of science.
Other researchers must be able to replicate the review process to confirm its reliability and accuracy. These issues are promoted in open science~\cite{mendezOpenScienceSoftware2020}, open data~\cite{runeson2024conceptual}, and in the principles of Findable, Accessible, Interoperable, and Reusable (FAIR) data endorsed in Nature in 2016~\cite{wilkinson2016fair}.

Reason 2: Trust.
Research artifacts build trust in the integrity of the evidence. When the methods, data, and process are fully available, readers and practitioners can trust that the conclusions are based on a thorough and unbiased synthesis of the evidence.

Reason 3: Updates.
Research artifacts are essential for updating SRs. Updating SRs ensures that the evidence remains current, preventing the research from becoming outdated or misleading~\cite{nepomucenoNeedUpdateSystematic2019}.

Reason 4: Pathway to Automation.
Writing an SR requires extensive time and effort -- on average, 67 weeks~\cite{marshallSystematicReviewAutomation2019}. Human errors may occur in the SR process due to reduced vigilance. Automation can help both in catching errors and reducing effort. Yet, research artifacts are needed to develop and validate the quality of automation. Recently, large language models (LLMs) have been used to automate SRs in SE by using existing research artifacts ~\cite{huotalaPromiseChallengesUsing2024, Felizardo2024ChatGPT}.
\section{Methods}
We conducted a systematic mapping study following the guidelines by Petersen et al.~\cite{petersenGuidelinesConductingSystematic2015a} and the SIGSOFT Empirical Standards checklist~\footnote{\url{https://www2.sigsoft.org/EmpiricalStandards/tools/}}. Full details of our research methods are available in our research artifact.~\footnote{\url{https://doi.org/10.5281/zenodo.15488074}\label{zenodoref}}.

\subsection{Search process}

The search was finalized on 2nd October 2024, using Scopus, which indexes metadata, including paper titles, abstracts, and keywords, from all reputable scientific publishers. The following search query was used: 
\begin{quote}
\footnotesize
\raggedright
\texttt{(ISSN(0928-8910) OR ISSN(1382-3256) OR ISSN(0950-5849) OR ISSN(2047-7481) OR ISSN(0164-1212) OR ISSN(0947-3602) OR ISSN(1619-1366) OR ISSN(0963-9314) OR ISSN(0960-0833) OR ISSN(0740-7459) OR ISSN(0038-0644) OR ISSN(1049-331x) OR ISSN(0098-5589) OR ISSN(0360-0300) OR ISSN(1876-7745) OR ISSN(1574-0137)) AND TITLE("Mapping Study" OR "Systematic review" OR "Systematic Literature Review" OR "Systematic Mapping" OR "Meta Analysis" OR "Meta Synthesis" OR "Scoping Review" OR "Case Survey" OR "Critical review") AND PUBYEAR > 2012 AND PUBYEAR < 2024}
\end{quote}

We searched for papers from 13 well-known SE-related journals, and from two reputable journals known for publishing reviews in the broader field of computer science. All selected journals are shown in Table~\ref{table:results_venues}.

We selected publications from 2013 to 2023, as  key data repositories such as Zenodo and Figshare were launched in 2013 and 2011, respectively. Our search used keywords related to systematic literature reviews, systematic mapping studies, meta-analysis and their corresponding variations, resulting in 643 articles.

\subsection{Study selection and quality assessment}
We applied the following inclusion criteria:
\begin{itemize}
\item IC1: Study published between 2013--2023
\item IC2: Study is a secondary study 
\item IC3: Study is related to SE
\end{itemize}

Title--abstract screening eliminated studies that were not secondary studies (IC2) or related to SE (IC3).
 
Research papers from ``ACM Computing Surveys" and ``Computer Science Review" journals were manually reviewed by the authors to select SE-related studies, as the two journals are not strictly limited to SE-related publications. After the manual review, the authors' inter-rater reliability was assessed using Krippendorff's Alpha~\footnote{\url{https://www.k-alpha.org/}}. The resulting score ($0.776$ on a $95\%$ confidence interval) indicates a strong agreement between the authors. After applying the inclusion criteria, 537 secondary studies remained.

\subsection{Data extraction}

Data extraction from the PDF files of secondary studies was conducted in two rounds: (1) manually and (2) automatically. In the manual phase, full-text screening is conducted for each included article to identify dedicated sections indicating the availability of research artifacts. In the automated phase, keyword search is conducted for the articles with a Python script that prints 100 characters before and after each keyword occurrence, which are then manually checked. We checked if the paper contains a reference to the external resource and if it is located in a permanent repository, such as Figshare, Zenodo or Mendeley.
\section{Results}\label{sec:results}
\begin{table*}
\caption{Publication channel and yearly statistics and the logistic regression model.}
\label{fig:results}
\centering
\begin{subtable}[t]{\textwidth}
	\caption{Comparison of the research artifact availability between publication channels}
        \label{table:results_venues}
	\centering
	\resizebox{0.9\linewidth}{!}{%
		\begin{tabular}{l|l|l:l|lll}
			Venue                                                    & Total & Yes          & Permanent repo & No           & By Request & Dead Link  \\
			\hline
			ACM Computing Surveys                                    & 29    & 8 (27.6\%)   & 4 (50.0\%)     & 20 (69.0\%)  & 0 (0\%)    & 1 (3.4\%)  \\
			ACM Transactions on Software Engineering and Methodology & 7     & 6 (85.7\%)   & 3 (50.0\%)     & 1 (14.3\%)   & 0 (0\%)    & 0 (0\%)    \\
			Automated Software Engineering                           & 1     & 1 (100.0\%)  & 0 (0\%)        & 0 (0\%)      & 0 (0\%)    & 0 (0\%)    \\
			Computer Science Review                                  & 14    & 1 (7.1\%)    & 0 (0\%)        & 12 (85.7\%)  & 0 (0\%)    & 1 (7.1\%)    \\
			Empirical Software Engineering                           & 24    & 12 (50.0\%)  & 4 (33.3\%)     & 12 (50.0\%)  & 0 (0\%)    & 0 (0\%)    \\
			IEEE Software                                            & 1     & 0 (0\%)      & 0 (0\%)        & 1 (100.0\%)  & 0 (0\%)    & 0 (0\%)    \\
			IEEE Transactions on Software Engineering                & 28    & 15 (53.6\%)  & 4 (26.7\%)     & 12 (42.9\%)  & 0 (0\%)    & 1 (3.6\%)  \\
			Information and Software Technology                      & 194   & 48 (24.7\%)  & 19 (39.6\%)    & 130 (67.0\%) & 8 (0.4\%)  & 8 (4.1\%)  \\
			Journal of Software: Evolution and Process               & 34    & 7 (20.6\%)   & 4 (57.1\%)     & 25 (73.5\%)  & 1 (2.9\%)  & 1 (2.9\%)  \\
			Journal of Systems and Software                          & 138   & 51 (37.0\%)  & 21 (41.2\%)    & 76 (55.1\%)  & 5 (3.6\%)  & 6 (4.3\%)  \\
			Requirements Engineering                                 & 15    & 4 (26.7\%)   & 2 (50.0\%)     & 9 (60.0\%)   & 0 (0\%)    & 2 (13.3\%) \\
			Software: Practice \& Experience                         & 12    & 1 (8.3\%)    & 0 (0\%)        & 10 (83.3\%)  & 1 (8.3\%)  & 0 (0\%)    \\
			Software and Systems Modeling                            & 23    & 10 (43.5\%)  & 2 (20.0\%)     & 11 (47.8\%)  & 1 (4.3\%)  & 1 (4.3\%)  \\
			Software Quality Journal                                 & 12    & 2 (16.7\%)   & 0 (0\%)        & 9 (75.0\%)   & 0 (0\%)    & 1 (8.3\%)  \\
			Software Testing: Verification \& Reliability            & 5     & 3 (60.0\%)   & 2 (66.7\%)     & 2 (40.0\%)   & 0 (0\%)    & 0 (0\%)    \\
			\hline
			Total                                                    & 537   & 169 (31.5\%) & 65 of 169 (38.5\%)    & 330 (61.5\%) & 16 (3.0\%) & 22 (4.1\%) \\
		\end{tabular}%
	}%
\end{subtable}
\hfill
\begin{subtable}[t]{\textwidth}
	\caption{Yearly statistics}
	\resizebox{\linewidth}{!}{%
		\begin{tabular}{l|rrrrrrrrrrr|l}
			Year                                                & 2013        & 2014        & 2015        & 2016        & 2017        & 2018        & 2019        & 2020        & 2021        & 2022        & 2023        & Total        \\
			\hline
			\multicolumn{5}{l}{Research artifact availability} & \multicolumn{7}{c|}{} & \\
			\hline
			Yes                                                 & 1 (3.7\%)   & 1 (4.2\%)   & 5 (10.4\%)  & 5 (16.1\%)  & 6 (14.3\%)  & 11 (16.7\%) & 15 (30.0\%) & 19 (38.8\%) & 22 (40.0\%) & 35 (53.0\%) & 49 (62.0\%) & 169 (31.5\%) \\
			No                                                  & 22 (81.5\%) & 20 (83.3\%) & 42 (87.5\%) & 22 (71.0\%) & 34 (81.0\%) & 53 (80.3\%) & 30 (60.0\%) & 28 (57.1\%) & 30 (54.5\%) & 28 (42.4\%) & 21 (26.6\%) & 330 (61.5\%) \\
			By req.                                             & 0 (0.0\%)   & 1 (4.2\%)   & 0 (0.0\%)   & 2 (6.5\%)   & 0 (0.0\%)   & 0 (0.0\%)   & 2 (4.0\%)   & 0 (0.0\%)   & 1 (1.8\%)   & 3 (4.5\%)   & 7 (8.9\%)   & 16 (3.0\%)   \\
			Dead                                                & 4 (14.8\%)  & 2 (8.3\%)   & 1 (2.1\%)   & 2 (6.5\%)   & 2 (4.8\%)   & 2 (3.0\%)   & 3 (6.0\%)   & 2 (4.1\%)   & 2 (3.6\%)   & 0 (0.0\%)   & 2 (2.5\%)   & 22 (4.1\%)   \\
			\hline
			\vtop{\hbox{\strut Permanent}\hbox{\strut repo}}    & 1 (3.7\%)   & 0 (0.0\%)   & 1 (2.1\%)   & 0 (0.0\%)   & 0 (0.0\%)   & 4 (6.1\%)   & 6 (12.0\%)  & 9 (18.4\%)  & 6 (10.9\%)  & 14 (21.2\%) & 24 (30.4\%) & 65 (12.1\%)  \\
			\hline
			\vtop{\hbox{\strut Dedicated}\hbox{\strut section}} & 0 (0.0\%)   & 0 (0.0\%)   & 0 (0.0\%)   & 0 (0.0\%)   & 2 (4.8\%)   & 2 (3.0\%)   & 1 (2.0\%)   & 5 (10.2\%)  & 3 (5.5\%)   & 13 (19.7\%) & 46 (58.2\%) & 72 (13.4\%)  \\
			\hline
			Total                                               & 27          & 24          & 48          & 31          & 42          & 66          & 50          & 49          & 55          & 66          & 79          & 537          \\
		\end{tabular}
	}%
\end{subtable}
\hfill
\begin{subtable}[t]{\textwidth}
\caption{Logistic regression model for the availability of research artifacts - year is a scaled ordered factor and comparison to a reference category (journal)}
\label{table:results_regression}
\centering
\resizebox{0.65\linewidth}{!}{%
\begin{tabular}{l|lrrrrr}
Coef & Est\textsuperscript{a} & Std. E & z value & P value & Odds ratio\\
\hline
Year(Ordered factor) &  0.84 &  0.12 &  7.21 & 5.79e-13$\textsuperscript{***}$ & 2.31 \\
\hline 
\multicolumn{5}{l}{Reference Category: IEEE Trans. Software Engineering} \\
\hline 
Empirical Software Engineering  &  -0.33  &  0.63 &  -0.52  & 0.94 &  0.61 \\
Computer Science Review &  -2.41  &  0.90 & -2.68 &  0.007** & 0.09\\
Software and Systems Modeling &  -0.79 &  0.62 & -1.27 &  0.20  & 0.46 \\
J. of Systems and Software  &  -0.61 &   0.48 & -1.26 &  0.21 & 0.55\\
J. Software: Evolution and Process &  -1.61  &  0.61 & -2.64 &  0.008** & 0.20\\
ACM Computing Surveys  &  -1.42 &   0.62 & -2.30 &  0.02* & 0.24\\
Software: Practice  Experience  &  -3.23  &  1.15& -2.81 &  0.005**  & 0.04\\
Requirements Engineering  &  -0.94  &  0.71 & -1.32 &  0.60  &  0.19\\
Information and Software Tech  &  -1.00 &   0.48 & -2.12  & 0.03*  & 0.37 \\
Software Quality J.  &  -1.13  &  0.81 & -1.39  & 0.17  & 0.17 \\
\hline
\end{tabular}
}%
\end{subtable}
\end{table*}

\def\firstrq{What proportion of secondary studies include a research artifact?}
\def\secondrq{Where are the research artifacts stored?}
\def\thirdrq{How is the availability of data stated?}
\def\fourthrq{How does publication year and the publication forum impact the availability of research artifacts?}

The results are shown in Table \ref{fig:results}. 

\textbf{RQ1: \firstrq}  
RQ1 gives an overview of the availability of research artifacts in SE SRs. Ideally, every secondary study would include a research artifact accessible through a direct link. 
Results: In our dataset, 169 of the 537 (31.5\%) reviewed secondary studies include a research artifact.

\textbf{RQ2: \secondrq} 
RQ2 addresses the storage location for research artifacts. Research artifacts should be located in permanent storage with a digital object identifier (DOI). Links to personal or institutional web pages may stop working in a few years.
The permanent repository (Zenodo, Figshare, or Mendeley Data) is linked in 38.5\% (65 out of 169 studies) of the studies that include a research artifact. The overall percentage of studies with permanent repository is only 12.1\% out of all papers (65 of 537).

\textbf{RQ3: \thirdrq} 
RQ3 focuses on how the availability of research artifacts is reported in research papers. 
Out of the studies that include a research artifact, 50 of 169 (29.6\%) include a dedicated section that states the availability of data or a research artifact.

\textbf{RQ4: \fourthrq}

Table \ref{table:results_regression} shows a binary logistic regression model, which models whether a research artifact is available, based on the publication year and journal. Journals with less than 10 publications in our dataset are excluded from this analysis. We used ``IEEE Transactions on Software Engineering" as the reference category due to its esteemed reputation in the field of SE. 

Publication year, modeled and scaled as an ordered factor for interpretability, is a significant predictor of whether a paper includes a research artifact. The model shows that for every 3 years (one standard deviation), the odds of the publication having a research artifact increase by a factor of 2.31.

All coefficients for journals are negative, meaning they are less likely to contain a research artifact than the reference category. Additionally, this difference is statistically significant for articles published in ``Computer Science Review", ``Software: Practice and Experience", ``Journal of Software: Evolution and Process" and ``Information and Software Technology", which are less likely to include a research artifact, with odds ratios ranging between 0.04 and 0.37.
\section{Limitations}

We excluded conference proceedings for two main reasons. First, most high-quality secondary studies are published in journals. Second, conference proceedings often lack ISSN numbers, and their quality can vary from year to year, with some years including colocated workshop papers and others not. While this decision may have caused us to miss some secondary studies, it is unlikely to have affected the conclusions of this study.

We chose not to use multiple academic databases, as Scopus already includes metadata from all relevant databases. However, if our search had been focused on full-text content, using multiple academic databases would have been mandatory.

Only secondary studies published between 2013--2023 were included.
By narrowing down the publication year, we wanted to make sure that permanent repository providers, such as Zenodo and Figshare, are fully included.

\section{Conclusion and Future work}
Our regression model indicates that secondary studies in software engineering are increasingly adopting open data practices.
In 2023, approximately 62.0\% of secondary studies include a research artifact, a figure we hope will approach 100\% in the future. Only 38.5\% of studies with research artifacts use permanent repositories with DOIs, which represents another area for improvement. Fortunately, we are also seeing progress in this area, with the overall average for permanent repositories being just 12.1\% in all SRs, while in 2023 it had already risen to 30.4\%. 

The importance of permanent repositories is highlighted, as links to non-permanent repositories can become inaccessible. In our study, even among papers published in 2023, 2 out of 19 links to non-permanent repositories are already dead.

Regarding 2023, the best adoption is in the papers' structure itself; 58.2\% of all papers contain a dedicated section stating the availability of data in 2023. Yet, we found that some papers with their "Data Availability" section simply state that ``no data was used" or something to a similar effect. Some papers mention that data is only ``available upon request". Both of these approaches are alarming because secondary studies rely only on published academic studies, and should not encounter confidentiality limitations.

In the future, journals should enforce the reporting practices of research artifacts to increase the availability of data. Additionally, identifying which research artifacts covered in this study have sufficient quality is an important future study area. 

Currently, our recommendation to all entities publishing secondary studies is to have a data availability section as it makes locating the research artifact easier. Additionally, research artifacts should be stored in a permanent repository with DOIs to ensure long term accessibility.
\section*{CRediT author statement}

\textbf{Aleksi Huotala:} Writing -- review \& editing, Writing -- Original Draft, Conceptualization, Data Curation, Methodology. \textbf{Miikka Kuutila:} Writing -- Review \& Editing, Writing -- Original Draft, Conceptualization, Formal Analysis, Methodology. \textbf{Mika Mäntylä:} Writing -- Review \& Editing, Writing -- Original Draft, Conceptualization, Methodology.

\section*{Declaration of competing interest}

The authors declare that they have no known competing financial interests or personal relationships that could have appeared to influence the work reported in this paper.

\section*{Acknowledgements}

The first author has been funded by the Strategic Research Council of Research Council of Finland (Grant ID 358471)
and the second author has been funded by the Killam Postdoctoral Fellowship.

\section*{Data availability}

The data used to conduct the research is available in Zenodo~\footref{zenodoref}

\bibliographystyle{elsarticle-num}
\bibliography{sources.bib}

\begin{thebibliography}{10}
\expandafter\ifx\csname url\endcsname\relax
  \def\url#1{\texttt{#1}}\fi
\expandafter\ifx\csname urlprefix\endcsname\relax\def\urlprefix{URL }\fi
\expandafter\ifx\csname href\endcsname\relax
  \def\href#1#2{#2} \def\path#1{#1}\fi

\bibitem{cruz2019replication}
M.~Cruz, B.~Bern{\'a}rdez, A.~Dur{\'a}n, J.~A. Galindo, A.~Ruiz-Cort{\'e}s, Replication of studies in empirical software engineering: A systematic mapping study, from 2013 to 2018, IEEE Access 8 (2019) 26773--26791.

\bibitem{heumuller2020publish}
R.~Heum{\"u}ller, S.~Nielebock, J.~Kr{\"u}ger, F.~Ortmeier, Publish or perish, but do not forget your software artifacts, Empirical Software Engineering 25~(6) (2020) 4585--4616.

\bibitem{mendezOpenScienceSoftware2020}
D.~Mendez, D.~Graziotin, S.~Wagner, H.~Seibold, Open {{Science}} in {{Software Engineering}}, in: M.~Felderer, G.~H. Travassos (Eds.), Contemporary {{Empirical Methods}} in {{Software Engineering}}, Springer International Publishing, Cham, 2020, pp. 477--501.
\newblock \href {https://doi.org/10.1007/978-3-030-32489-6_17} {\path{doi:10.1007/978-3-030-32489-6_17}}.

\bibitem{runeson2024conceptual}
P.~Runeson, E.~Soderberg, M.~Host, A conceptual framework and recommendations for open data and artifacts in empirical software engineering, in: Proceedings of the 1st IEEE/ACM International Workshop on Methodological Issues with Empirical Studies in Software Engineering, 2024, pp. 68--75.

\bibitem{wilkinson2016fair}
M.~D. Wilkinson, M.~Dumontier, I.~J. Aalbersberg, G.~Appleton, M.~Axton, A.~Baak, N.~Blomberg, J.-W. Boiten, L.~B. da~Silva~Santos, P.~E. Bourne, et~al., The fair guiding principles for scientific data management and stewardship, Scientific data 3~(1) (2016) 1--9.

\bibitem{nepomucenoNeedUpdateSystematic2019}
V.~Nepomuceno, S.~Soares, On the need to update systematic literature reviews, Information and Software Technology 109 (2019) 40--42.
\newblock \href {https://doi.org/10.1016/j.infsof.2019.01.005} {\path{doi:10.1016/j.infsof.2019.01.005}}.

\bibitem{marshallSystematicReviewAutomation2019}
I.~J. Marshall, B.~C. Wallace, Toward systematic review automation: a practical guide to using machine learning tools in research synthesis, Systematic Reviews 8~(1) (2019) 163.
\newblock \href {https://doi.org/10.1186/s13643-019-1074-9} {\path{doi:10.1186/s13643-019-1074-9}}.

\bibitem{huotalaPromiseChallengesUsing2024}
A.~Huotala, M.~Kuutila, P.~Ralph, M.~M\"{a}ntyl\"{a}, The promise and challenges of using llms to accelerate the screening process of systematic reviews, in: Proceedings of the 28th International Conference on Evaluation and Assessment in Software Engineering, EASE '24, Association for Computing Machinery, New York, NY, USA, 2024, p. 262–271.
\newblock \href {https://doi.org/10.1145/3661167.3661172} {\path{doi:10.1145/3661167.3661172}}.

\bibitem{Felizardo2024ChatGPT}
K.~Felizardo, M.~Lima, A.~Deizepe, T.~Conte, I.~Steinmacher, Chatgpt application in systematic literature reviews in software engineering: an evaluation of its accuracy to support the selection activity, in: Proceedings of the 18th ACM/IEEE International Symposium on Empirical Software Engineering and Measurement, 2024, pp. 1--10.

\bibitem{petersenGuidelinesConductingSystematic2015a}
K.~Petersen, S.~Vakkalanka, L.~Kuzniarz, Guidelines for conducting systematic mapping studies in software engineering: {{An}} update, Information and Software Technology 64 (2015) 1--18.
\newblock \href {https://doi.org/10.1016/j.infsof.2015.03.007} {\path{doi:10.1016/j.infsof.2015.03.007}}.

\end{thebibliography}
\end{document}